\documentstyle[preprint,aps]{revtex}
\tighten
\draft
\widetext
\preprint{CLNS 98/1561}
\bigskip
\bigskip
\begin{document}
\title{TeV Scale Superstring and Extra Dimensions}
\medskip
\author{Gary Shiu\footnote{E-mail: shiu@mail.lns.cornell.edu} 
and S.-H. Henry
Tye\footnote{E-mail: tye@mail.lns.cornell.edu}}
\bigskip
\address{Newman Laboratory of Nuclear Studies, Cornell University,
Ithaca, NY 14853}
\date{May 24, 1998}
\bigskip
\medskip
\maketitle

\begin{abstract}
{}Utilizing the idea of extra large dimensions, it has been suggested that 
the gauge and gravity couplings unification can happen at a scale as low as 
1 TeV. In this paper, we explore this phenomenological
possibility within string theory.
In particular, we discuss how the proton decay bound can be satisfied in 
Type I string theory. 
The string picture also suggests different scenarios of gauge and
gravitational couplings unification. The various scenarios are explicitly 
illustrated with a specific $4$-dimensional ${\cal N}=1$ supersymmetric 
chiral Type I string model with Pati-Salam-like gauge symmetry. 
We point out certain features that should be generic in other Type I strings.

\end{abstract}

\section{Introduction}

{}Probably the most important problem in elementary particle theory today is to
find out how superstring theory describes our universe. In the standard
scenario, the Planck scale $M_P$ ({\em i.e.}, $10^{19}$ GeV) defines 
the string scale to be around $10^{17}$ GeV, while the unification of 
the gauge couplings happens at the grand unified scale $M_{GUT}$ 
(around $10^{16}$ GeV) \cite{gut}. Although the string scale and 
$M_{GUT}$ are quite close, 
the discrepancy between them may still be of some concern. However, 
a more practical problem with this scenario is the difficulty
in calculating physical observables. Since the natural string scale of 
this scenario is between $M_{GUT}$ and $M_P$, 
while most of the physical observables are at the electroweak scale 
$M_{EW}$, a typical  comparison between theory and experiment requires a 
detailed analysis of a specific string model. Unfortunately, our understanding
of the string dynamics is still quite primitive, making such precise
calculations essentially impossible. So the connection between string theory
and our observable universe is rather tenuous in this scenario at this moment.
It is therefore exciting that an alternative scenario has recently emerged.

{}The idea of extra dimensions have been well studied in Kaluza-Klein 
theories and string theories. 
It was suggested by Antoniadis {\em et. al.} \cite{Antoniadis} 
that, beyond the usual 4 space-time dimensions that we live in, there are 
large extra dimensions that may be probed by upcoming experiments.
It is by now well-known that, in some string models, the gravity lives
in the bulk, while the gauge and charged 
matter fields live on the branes (which
may be understood as special types of solitons and can have lower
dimensions than the bulk \cite{Dai}); our $4$-dimensional universe may
actually be inside the branes \cite{witt}. 
In particular, the extra dimensions that gravity feels can be
as large as 1 mm, as recently pointed out by Arkani-Hamed, Dimopoulos and 
Dvali \cite{extra}, while the extra dimensions that the gauge and other 
matter couplings feel can be as big as $M_{EW}^{-1}$, 
as recently pointed out by Dienes, Dudas and Gherghetta \cite{dienes}. 
In this scenario, 
the Planck scale $M_P$ is traded for the size of the extra dimensions
felt by gravity \cite{extra,extraI}. Likewise,
gauge coupling unification can be preserved and remain perturbative,
but now occurs at scales as low as a TeV \cite{dienes}.
One can therefore now have gravity and gauge coupling unification 
as low as a few hundred GeV to 1 TeV.
In string theory, this means that the string scale $m_s$
can be as low as a TeV. 
Such a scenario has been suggested previously by Lykken \cite{lykken}.
One advantage of this TeV scale string scenario is obvious. 
Not only that near future experiments can probe the string scale
and the large extra dimensions, 
it may even help us unravel the string
dynamics and pinpoint the string vacuum we live in.

{}In addition to the advantage of being experimentally testable, this new 
scenario may offer a simple qualitative explanation to the fermion mass 
hierarchy problem, as pointed out in Ref \cite{dienes}.
To be specific, let us suppose the string scale $m_s=$ 1 TeV.
Gravity, but not the standard model gauge and matter fields, lives in
$n$ large compactified dimensions, with radii $r_i$.
The radii $R_j$ of the remaining compactified dimensions in which 
both gravity and gauge fields live are
somewhere between $m_{s}^{-1}$ and $M_{EW}^{-1}$, so  
$m_s > R_j^{-1} > M_{EW} >> r_i^{-1}$.
In this scenario, the effective couplings at 
the $m_s$ scale are all irrelevant operators and so the dimensionless
gauge couplings $\alpha_i$ and Yukawa couplings $y_f$
run as powers of the energy scale. If the different Yukawa 
couplings are comparable at the $m_s$ scale, 
they can easily differ by orders of magnitude at
the electroweak scale due to this power-law behavior.
The gauge couplings differ by only one order of magnitude because they are
unified at the string scale. Below the $R_j^{-1}$ scales,
the dimensionless gravitational coupling runs a function of the energy scale 
like $E^{2+n}$ to the scales $r_i^{-1}$
and then runs like $E^2$, yielding a huge $M_P$ \cite{witt}. So the 
presence of the extra dimensions provide a qualitative explanation of 
the origin of the orders of magnitude differences among the couplings.
As pointed out in Ref \cite{dienes}, $m_s >>$ 1 TeV is perfectly acceptable.
However, we have to treat $m_s$ and $M_{EW}$ as two different scales in this 
situation.

{} In this paper, we study a number of issues in the $4$-dimensional
${\cal N}=1$ 
chiral Type I string theory, which is the appropriate framework for the
TeV scale string scenario. A typical model will have $9$-branes, which fill 
the $10$-dimensional spacetime, and $5$-branes, which fill $6$-dimensional 
spacetime. Both branes have a flat $4$-dimensional uncompactified spacetime.
The string picture has been discussed previously in Ref \cite{lykken,extraI}. 
Here, we review and extend the analysis.
Among other observations in this paper, we note that:

$\bullet$ It is well-known that proton decay can be suppressed by 
symmetry. Here, we see that proton decay is suppressed by 
the presence of a custodial $U(1)$ gauge symmetry. 
The presence of such a $U(1)$ gauge symmetry is generic in Type I strings.

$\bullet$ As an alternative scenario, the standard model gauge symmetries can 
come from different types of branes, e.g., QCD $SU(3)$ comes from one 
type of branes (say, 9-branes) while the weak $SU(2)$ comes from 
another type (say, 5-branes). Since the $9$-brane couplings are in general 
different from the $5$-brane couplings, the standard model gauge 
couplings do not need to meet at the string scale. Rather, an 
appropriate choice of the sizes of the compactified dimensions is needed for
the couplings to agree with experiment.

$\bullet$ Cavendish type experiments have tested Newton's Law to a scale 
of millimeters \cite{price}, providing an upper bound on the large radius. 
The strong and electroweak scatterings have tested the small extra dimensions
to a radius of $M_{EW}^{-1}$, providing an upper bound on the size of the 
small extra dimensions. Taking $m_s R_j \sim 1$,
the relation between the large radii $r_i$ and $M_P$ is given by
\begin{equation}
M_P^2 \sim  32 \pi^2 g^{-4} m_s^2 \prod_{i=1}^{n} (m_s r_i) 
%\prod_{j=1}^{6-n} (m_s R_j) 
%r^n = {M_P^2 \over {16 (2 \pi)^{8+n}}} m_s^{-(2+n)}
\end{equation}
where $n$ is the number of large compactified dimensions, $g$ is the gauge 
coupling \cite{extraI}. 
The numerical factor follows from string unitarity and duality.
Clearly $m_s$ must be bigger than $M_{EW}$. 
Assume the gauge coupling $g^2 \sim 1$.
For $n=2$, $r$ is about $10^{-4}$ meter for $m_s$= 1 TeV. 
As pointed out in 
Ref \cite{extra}, both the 1 mm scale and the 1 TeV scale can be tested by 
experiments in the near future. We point out that the $n=2$ choice seems 
natural in a number of string scenarios.
For example, in the specific model that we consider in this paper, 
only the $n=2$
choice gives rise to 3 chiral families.

$\bullet$ String theory has no global symmetry. However, some gauge 
couplings are proportional to $r^{-1}$. For very large r, they become 
so weak that the respective gauge symmetries may appear like 
global symmetries. In some situations, the corresponding matter fields 
with vanishingly small gauge couplings are suitable candidates for dark matter.

{}To make the discussion concrete, we construct an explicit model to
illustrate the TeV scale string scenario. Our analysis of the model is 
quite sketchy and cavalier. Our purpose is to draw attention to the model's 
features that are generic to other Type I string models.
The model is a $D=4$, ${\cal N}=1$ supersymmetric, 
chiral Type I string model, with $9$-branes and $5$-branes. Their gauge 
groups $G_9$ and $G_5$ (with gauge couplings $g_9$ and $g_5$ respectively)
are identical:
$G_9$=$G_5$= $U(4) \otimes U(2) \otimes U(2)^{\prime}$. 
The massless open string spectrum is given in Table \ref{Z6}. 
The $U(1)$'s associated with 
the $SU(2)$s are anomalous, not unusual in string theories. 
The $U(1)$ associated with the $SU(4)$ provides the custodial symmetry to 
suppress proton decay.
We shall use this model to discuss the following three scenarios:

$\bullet$ One may identify the $SU(4) \otimes SU(2) \otimes SU(2)^{\prime}$ 
from the $9$-brane sector as the Pati-Salam group. Spontaneous symmetry 
breaking reduces it to $SU(3) \otimes SU(2)_L \otimes U(1)$. 
At the string scale, both the QCD coupling $g_3$ and the weak coupling 
$g_2$ are equal to the $9$-brane coupling $g_9$, and $\sin^2 \theta_W=3/8$. 
There is a $U(1)_B$ gauge symmetry 
associated with the baryon number, so the proton decay is suppressed. 
However, it seems that breaking this $U(1)_B$ symmetry will also 
break QCD $SU(3)$. The model has only one chiral family, plus a vector 
({\em i.e.}, a chiral and an anti-chiral) family.

$\bullet$ One may identify 
$SU(4)_9 \otimes SU(2)_5 \otimes SU(2)_9$ (the subscripts
indicate which sectors each comes from) as the Pati-Salam gauge group. 
The QCD $SU(3)$ comes from the spontaneous symmetry breaking of the 
$SU(4)_9 \otimes SU(2)^{\prime}_5$ by
a bi-fundamental matter field, while the remaining $SU(2)_5$ is identified with
the weak $SU(2)_L$. The $SU(4)_5$ may get strong and induce both dynamical
supersymmetry breaking and electroweak symmetry breaking. 
The gauge group $SU(2)_9 \otimes SU(2)^{\prime}_9$ is also
broken. At the string scale, $g_3=g_9$ while $g_2=g_5$.
Now there are two families of quarks and leptons, coming from the 95
sector, while the Higgs fields come from the 55 sector. Again, the perturbative
couplings obey the baryon quantum number conservation because of $U(1)_B$.  

$\bullet$ The first scenario has one chiral family in the 99 sector while 
the second scenario has two chiral families in the 59 sector. 
It turns out that there is another scenario where there are three chiral
families.
Under diagonal spontaneous symmetry breaking, $SU(2)_9 \otimes SU(2)_5$
becomes $SU(2)$. This Higgs mechanism is permitted by the presence of the 
appropriate bi-fundamental fields,
resulting in a model which contains $SU(4) \otimes SU(2)_L 
\otimes SU(2)_R$ with 3 chiral families:
one from the 99 sector and two from the 59 sector. 
Weak interaction universality automatically follows, independent of the 
relative values of $g_9$ and $g_5$. 
The standard model couplings are 
all different. In the case $g=g_9=g_5$, we have at the string scale,
\begin{equation}
 g^2 = g_3^2 = 2 g_2^2 = {8 \over 3} g_Y^2.
\end{equation}
%g_3=g, $g_2=g/\sqrt{2}$ and $g_1=g_Y=g/\sqrt{3}. 
Hence the electroweak Weinberg mixing angle satisfies 
$\sin^2 \theta_W=3/7$ at the string scale.
This scenario has one practical advantage. Since $g_3 > g_2 > g_Y$ at the 
string scale, the running couplings need power-like behavior for 
only a relatively short range of energies; that is, they do not have to 
grow much. As a consequence, the string coupling will stay weak, and 
perturbative Type I string theory should be valid for analysis.

{}It is clear that, among other properties, the presence of $U(1)$'s 
(associated with the centers of mass of the branes), the presence of 
bi-fundamental matter fields, and the identical nature of $9$-brane 
gauge group and $5$-brane gauge groups (if present) are generic 
features of many $D=4$, ${\cal N}=1$ 
supersymmetric, chiral Type I string models. These properties are quite 
compatible with experiments and the extra large dimension scenario. Further 
investigation along this direction will certainly be worthwhile. 

{}This paper is organized as follows. In Section 2, the basic idea of the
TeV scale string scenario is reviewed. 
The Type I string picture with extra large dimensions is discussed.
As an illustration, the Pati-Salam like Type I string model is presented 
in Section 3. 
Among other issues, we see how the proton decay bound can be resolved.
Section 4 contains some discussions on the various issues in the scenario.
Section 5 contains the comments. The details of the
construction of the Type I string model is contained in Appendix A. 
It is relegated to an appendix because of its technical nature.  
In Appendix B, we review how to calculate amplitudes to determine the terms 
in the superpotential of the model.

\section{Brane Picture}

The idea of extra large dimensions is most conveniently realized in terms
of Type I string theory and D-branes \cite{Dbranes}. The graviton (coming 
from the closed string sector) lives in the bulk, while the gauge and 
charged matter fields (coming from the open string sector)
live on the branes (which is $p+1$-dimensional for a $Dp$-brane). 
Since gravity and gauge fields see different numbers
of dimensions, it is possible to have extra large dimensions without
making the gauge couplings extremely small at low energies.
In the worldsheet construction of heterotic string model,
both gravity and gauge fields live in the same space, and so
the idea of extra dimensions is difficult to implement.
It is possible to realize the extra dimensions with the 
solitonic 5-branes in heterotic string theory.
However, the techniques in constructing heterotic string model
with these solitonic 5-branes are not very well developed.
As a result, Type I string theory and D-branes provide the most natural
setting to understand the generic features of extra dimensions.
Here, we review and expand on the earlier 
discussions \cite{lykken,extraI}.

\subsection{Supersymmetric Type I String}

{}Various ${\cal N}=1$, $D=4$
Type I string models have been studied in the past two years 
\cite{BL,Sagnotti,ZK,KS1,KS2,Zw,Iba,long}. 
They are especially suitable for realizing the idea of extra large dimensions.
Gravity lives in the bulk while the gauge fields live on the branes. There are
9-branes, which overlap completely with the bulk. 
If a model has both $p$-branes and $q$-branes, then supersymmetry imposes 
the restriction $p-q=0$ (mod 4). To keep the Lorentz property of 4-dimensional
spacetime, only 5-branes and 9-branes
are permissible. So, for some models, there can be 
5-branes as well. 

{}The 4-dimensional string has the usual 4 spacetime dimensions 
($x_0,\dots,x_3$), and 6 compactified dimensions. We shall 
treat this 6 dimensions $T^6$ as composed of 3 two-tori: 
$T_1$ (with coordinates 
$x_8$, $x_9$), $T_2$ (with coordinates $x_6$, $x_7$) 
and $T_3$ (with coordinates $x_4$, $x_5$), 
the volumes of which are $v_1$, $v_2$ and $v_3$ 
respectively. So the volume of the 6 compactified 
dimensions is $v_1 v_2 v_3$. Crudely speaking, the volume $v_i$ 
can be expressed 
in terms of the 
compactified radius $r_i$, $v_i= (2\pi r_i)^2$. \footnote{The radius $r_i$
here does not necessarily have to be the radius $R_i$ of the torus. It is
simply a characteristic length scale of the compactified dimension.
In the case of a ${\bf Z}_N$ orbifold, the volume is given 
by $\prod_i v_i={1\over N} \prod_{i} (2 \pi R_i)^2  \equiv \prod_{i} 
(2 \pi r_i)^2$.} 
So the low energy effective action is given by
\begin{equation}\label{95action}
S = \int d^4 x \sqrt{g} \left( 
{{m_s^8 v_1 v_2 v_3} \over (2 \pi)^7 \lambda^2} R 
+ {1\over 4} {{m_s^6 v_1 v_2 v_3} \over {(2 \pi)^7 \lambda}} F^2
+ {1 \over 4}
\sum_{i=1}^{3} {{m_s^2 v_i} \over {(2 \pi)^3 \lambda}} 
\tilde{F}_i^2 +\dots \right)
\end{equation}
where $m_s$ is the string scale. $F$ is the field strength of the gauge fields 
in the 9-branes while the $\tilde{F}_i$ is 
the gauge field strength on different types of 5-branes 
(the worldvolumes of which are ${\cal M}^4 \times T_1$,
${\cal M}^4 \times T_2$ and ${\cal M}^4 \times T_3$ respectively;
${\cal M}^4$ being the 4 dimensional Minkowski space-time).
Here $\lambda$ is the string 
coupling, {\em i.e.}, $\lambda \sim e^{\phi}$, where 
$\phi$ is the dilaton field.
%The factor of $2 (2 \pi)^7$ in the Hilbert-Einstein term 
%comes from the non-trivial relation between the open string
%and closed string loop expansion parameters \cite{sakai}. 
The relative normalization of the Newton's constant and the gauge
coupling (which is related to the D-brane tension) is
obtained by factorizing scattering amplitudes
into open and closed string channels \cite{Dbranes,sakai}. 
(In Type I string, the $N$-point open-string one-loop amplitude is 
equivalent to the closed string scattering to $N$ open strings 
at the tree level. This relation follows from unitarity).
This should be compared to heterotic
string theory where all states are closed string states.
The precise numerical factors are determined once we define
the string coupling $\lambda$ to be the ratio of the
fundamental string and D-string tensions in Type IIB string 
theory \cite{Dbranes}.

{}For simplicity, we will consider only one type of 5-branes in what 
follows.
Let $G_9$ ($G_5$) be the gauge group of the 9-brane (5-brane). 
The 5-branes are compactified on $T_3$, 
while the 9-branes are compactified on  
$T^6$. 
The branes and the bulk have a common 4-dimensional 
uncompactified spacetime.
The 4-dimensional Planck mass $M_P$ and the
Newton's constant $G_N$ are given by
\begin{equation}\label{newton}
G_N^{-1}=M_P^2 = 
{{8 m_s^8 v_1 v_2 v_3} \over (2 \pi)^6 \lambda^2}
\end{equation}
and the gauge couplings of $G_9$ and $G_5$ are
\begin{equation}
g_9^{-2} = {{m_s^6 v_1 v_2 v_3} \over (2 \pi)^7 \lambda} ~, \quad \quad
g_5^{-2} = {{m_s^2 v_3} \over (2 \pi)^3 \lambda}
\end{equation}
These relations are subject to quantum corrections, which we shall ignore 
for the moment.
{}Recall that the gauge couplings of the standard model are of order 1. 
In string theory, there is a T-duality symmetry, {\em i.e.}, physics is 
invariant 
under a T-duality transformation.
If any of the volume $v_i$ is much smaller than the string scale, {\em i.e.}, 
$v_i$ less than $m_s^{-2}$, the T-dual description is more convenient:
\begin{eqnarray}
	 \lambda &\rightarrow& (2 \pi)^2
         {\lambda \over {v_i m_s^2}}  \nonumber \\
  	 v_i &\rightarrow & (2 \pi)^4 {1 \over {v_i m_s^4}}
\end{eqnarray}

{}In this dual picture, the new volume $(2 \pi)^4/(v_i m_s^4)$ 
of the dual $T_i$ torus is 
large. Under this duality transformation, the Dirichlet and Neumann
boundary conditions of the open strings are interchanged, and so 
the branes are also 
mapped to other types of branes. For example, for $i=1$, {\em i.e.}, we T-dual 
the $T_1$ torus; the 9-branes become 7-branes ($x_0,\dots,x_7$), 
while the 5-branes become 7-branes ($x_0,\dots,x_5,x_8,x_9$). 
Therefore, they are orthogonal in the 
compactified space. The effective action becomes
\begin{equation}
S = \int d^4 x \sqrt{g} \left( 
{{m_s^8 v_1 v_2 v_3} \over (2 \pi)^7 \lambda^2} R 
+ {1 \over 4} {{m_s^4 v_2 v_3} \over {(2 \pi)^5 \lambda}} F^2
+ {1 \over 4} {{m_s^4 v_1 v_3} \over {(2 \pi)^5 \lambda}} 
\tilde{F}^2 +\dots \right)
\end{equation}

{}If the standard model gauge group is in $G_9$, then, in 
this $7$-brane picture,
\begin{equation}
g_9^{-2} = {{m_s^4 v_2 v_3} \over (2 \pi)^5 \lambda} \sim 1
\end{equation}

{}Now, suppose $m_s$ is 1 TeV. To satisfy Eq.(\ref{newton}), 
{\em i.e.}, to obtain the $M_P=10^{19}$ GeV, at least one of the 
2-volumes must be large. Since the $G_9$ gauge coupling must be of
order 1, the 
only choice is to take $v_1$ large. This means the $G_5$ gauge coupling becomes
extremely small, {\em i.e.}, the gauge fields decouple. The conserved currents 
that couple to $G_5$ will appear like those of global symmetries.

{}We can also keep both $G_9$ and $G_5$ gauge couplings of order 1.
This can be achieved if we T-dual both the $T_1$ and $T_3$ tori to end up with 
orthogonal (in the compactified space) 5-branes with the following 
effective action:
\begin{equation}\label{55action}
S = \int d^4 x \sqrt{g} \left( 
{{m^8 v_1 v_2 v_3} \over (2 \pi)^7 \lambda^2} R 
+ {1 \over 4} {{m_s^2 v_2} \over {(2 \pi)^3 \lambda}}  F^2
+ {1 \over 4} {{m_s^2 v_1} \over {(2 \pi)^3 \lambda}}  
\tilde{F}^2 +\dots \right)
\end{equation}
In this case, we can take $v_3$ large to satisfy Eq.(\ref{newton}).
We shall use this $55'$-brane picture later, but we shall still 
refer to the $5$-branes coming from T-dualizing
the $9$-branes as the $9$-branes.
To summarize, there are 8 inequivalent scenarios one can entertain: 95-, 77-,
55- and 73-brane configuration, and their $T^6$-duals ({\em i.e.},
T-dual along all 6 dimensions): 59-, 77-, 55- and 
37-brane configurations. The 95-, 77- and 55-brane effective actions are 
given above. 

\noindent
$\bullet$ We have kept the two radii in each torus to be the same. To build 
an ${\cal N}=1$ 
supersymmetric model, we need to orbifold the compactified dimensions
in the complexified basis. Equal radii in each torus yield discrete 
symmetries that can be gauged in orbifolds. 

\noindent
$\bullet$ If $G_9$ is identical to $G_5$, and the matter fields and 
couplings are 
symmetric under the interchange of 9- and 5- sectors, then the above 8 cases 
reduce to 4 inequivalent cases. This seems to be the generic situation in 
simple Type I model-building.  

\noindent
$\bullet$ To keep at least one sector of gauge fields visible ({\em i.e.}, 
gauge 
coupling of order 1), we can take at most two $T^2$'s, say $T_1$ and $T_3$, 
with large radii. 
Eq.(\ref{55action}) implies that the product of the two radii is
\begin{equation}
M_P^2 = {{8 m_s^8 v_1 v_2 v_3} \over {(2 \pi)^6\lambda^2}} 
\sim 32 \pi^2 m_s^6 r_2^2 r_3^2
\end{equation}
If they are equal, then the radius is around $10^{-12}$m for $m_s=1$ TeV.

\noindent
$\bullet$ If we want both $G_9$ and $G_5$ to be observable, we can 
take only one 
$T^2$ to have large radius. In the effective action (\ref{55action}), 
we can take $v_3$ large.
This scenario is necessary in any one of the following situations: 

	(i) the standard model is contained in one sector, say $G_9$, while 
a large gauge coupling from $G_5$ may be needed for a strong interaction to 
generate dynamical supersymmetry breaking. 

	(ii) the standard model is contained in both $G_9$ and $G_5$, 
(for example QCD $SU(3)$ in $G_9$ while weak $SU(2)_L$ in $G_5$). 
In this case, the $G_9$ and $G_5$ gauge couplings are in general 
different even at the string scale:
\begin{equation}\label{gaugecoupling}
g_3^{-2}=g_9^{-2}= {{m_s^2 v_2} \over (2 \pi)^3 \lambda}, \quad \quad
g_2^{-2}=g_5^{-2}= {{m_s^2 v_1} \over (2 \pi)^3 \lambda}
\end{equation}

	(iii) QCD $SU(3)$ is inside the $9$-branes while the weak $SU(2)_L$
comes from the diagonal spontaneous symmetry breaking of a $SU(2)$ 
inside the $9$-branes and a $SU(2)$
inside the $5$-branes. In this case, the standard model gauge couplings 
$g_3$, $g_2$ and $g_1$ are in general different at the string scale, even 
if $g_9=g_5$. 

We shall illustrate each of these possibilities in the next section. From 
equation (\ref{newton}),
\begin{equation}
r_3 \sim 10^{-4} g_{9} g_{5} \left({m_s \over {\mbox TeV}} \right)^{-2} 
\mbox{meter}
\end{equation}
If both $g_5$ and $g_9$ are of order unity, and $m_s$ is 1 TeV, then $r$ is
$10^{-4}$ meter.
If $g_5$ becomes small (equivalent to large radius $r_1$), 
$r_3$
just becomes even smaller. 

{}Let us go back to the general case with three types of 5-branes 
(as in Eq.(\ref{95action})). 
%A 4-dimensional chiral ${\cal N}=1$ Type I string model of this 
%type is given in Ref\cite{zura6}. 
Similar analysis is easy to carry out, so we shall simply
restrict ourselves to a few comments. It is easy to see that under
T-duality, the rule $p-p'=0$ (mod 4) is preserved. If $v_3$ gets large, we
see that the gauge couplings of both the 9-brane gauge sector and the
third type of 5-brane gauge sector become vanishingly small. As a
consequence, the matter fields in this particular 59 sector will
essentially decouple from all gauge interactions. They will still couple
to other fields via other interactions, including gravity. So they are
suitable candidates for dark matter.

\subsection{Non-supersymmetric String and the Cosmological Constant}

{}Supersymmetry was introduced originally to solve the hierarchy problem. 
Since this hierarchy problem disappears when the string scale is close to 
the weak scale, we should also consider non-supersymmetric Type I models. 
Generically, besides 9-branes, 7-, 5- and/or 3-branes may be present in 
a specific model, depending on the details. 
%So the effective action looks like
%\begin{equation}
%S= \int d^4 x \sqrt{g} \left(   2 
%(2 \pi)^7 {{m_s^8 v_1 v_2 v_3} \over \lambda^2} R
%+ {{m_s^6 v_1 v_2 v_3} \over {4 \lambda}} F^2
%+ {{m_s^4 v_2 v_3} \over {4 \lambda}}  F_7^2
%+ {{m_s^2 v_3} \over {4 \lambda}} \tilde{F}^2
%+ {1 \over {4 \lambda}}  F_3^2 - \Lambda_4 \right) 
%+ \dots
%\end{equation}
We also expect a cosmological constant $\Lambda_4$ to be present.
Again, we can consider the inequivalent scenarios when the various tori 
become large. Besides the 9753-brane configuration, duality can bring us to 
the 7975-, 7575- and 7535-brane configurations.
Depending on the choice, taking one torus volume large will decouple gauge 
fields from one or more sectors (if they are present). The analysis is 
similar to that given for the supersymmetric case and will not be repeated 
here. There is one important difference between the supersymmetric case
and the non-supersymmetric case. For supersymmetry to be unbroken,
the 6 dimensional manifold must be a complex manifold. This means that
$T^6$ can always be written as $T^6=T^2 \otimes T^2 \otimes T^2$, 
where the two radii in each $T^2$ are the same (needed for orbifold symmetry).
In the non-supersymmetric case, it is possible that only one dimension has
large radius (this breaks the complex structure). 
%In this case, 
%\begin{equation}
%r \sim 10^{10} \left({m_s \over {\mbox{TeV}}} \right)^{-3} \mbox{meter}
%\end{equation}
%when both $g_9$ and $g_5$ are of order 1.
%Experiments may reach 1 mm. In this case, $m_s$ is around $10^4$ TeV.

{}Let us comment on the cosmological constant.
We have seen how a large Planck mass $M_P$ can be generated from a 
much smaller string scale $m_s$. Naively, the same effect happens to the 
cosmological constant. If there is a 10-dimensional 
cosmological constant $\Lambda_{10} = m_s^{10}$, then 
$\Lambda_4=\Lambda_{10} v_1 v_2 v_3$, which is obviously unacceptable. 
Fortunately, this argument is incorrect.
Recall the construction of the string model. We start from a 4-dimensional 
supersymmetric model toroidally compactified from 10 dimensions; it has no 
cosmological constant. We reduce the number of supersymmetries by
orbifolding/orientifolding.
The orbifolding of each of the three tori is needed to break the 
spacetime supersymmetry and generate chiral fermions, so the mechanism is 
intimately tied to $D=4$ spacetime. This suggests that 
$\Lambda_4 = m_s^4$. 
This is substantially smaller than the previous naive estimate.
Unfortunately, this is still unacceptably large, so we need to find 
some mechanism to suppress it further. Now that we have seen how extra 
large dimensions can blow up the Planck mass, we are naturally led to ask if 
the reverse can suppress the cosmological constant. 

	Suppose we construct a non-supersymmetric string model in 3 
spacetime dimensions. (In the construction of non-supersymmetric models,
we do not need to complexify the compactified dimensions.)
So generically $\Lambda_3 =m_s^3$. Now, let us take the radius $r$ of 
one of the compactified direction to be large, {\em i.e.}, 
decompactify that direction. So the theory essentially describes a 
$4$-dimensional spacetime. The 4-dimensional
cosmological constant is given by 
\begin{equation}
\Lambda_4 \sim {\Lambda_3 \over r} \sim {m_s^3 \over r}~. 
\end{equation}
For $m_s= 1$ TeV and $r$ the size of the 
universe, $\Lambda_4$ is small enough to be acceptable. This means 
the supersymmetry breaking mechanism within the string model-building must 
be intrinsically 3-dimensional. 
This imposes a strong constraint in non-supersymmetric string  model-building. 
Generically, the theory can decompactify in other directions in the 
field space, so that $\Lambda_4$ ends up of the order of $m_s^4$. However, 
$\Lambda_4$ measures the vacuum energy density, so it is natural for it to 
choose the minimum energy path of decompactification.  
This imposes a strong constraint in model-building. 
Notice that this mechanism will not work if the string scale 
is around the GUT scale, as is the case in the old scenario.

 	The above scenario is different from Witten's suggestion\cite{witten}, 
which also utilizes the 3 spacetime dimensional picture. In 3 dimensional
globally supersymmetric theories, the fermion-boson mass splitting 
$\delta m$ is zero, as naively expected, but becomes non-zero in 
supergravity models. This implies that $\delta m \sim m^2/M$, 
where $m$ is the 
typical mass and $M$ is the 3-dimensional Planck mass \cite{strominger}. 
So the fermion-boson mass splittings are non-zero while $\Lambda_3$ is zero.
As we decompactify a direction with radius $r$, $\Lambda_4$ clearly remains 
zero. However, the 4-dimensional $M_P^2=M/r$, so, for finite $M_P$, $M$ goes 
to infinity as $r$ goes to infinity, and $\delta m$ goes to zero. This seems 
to imply that the decompactification of the $3$-dimensional supergravity model
yields $4$-dimensional supergravity. 
So we believe that 
non-supersymmetric 4-dimensional models can come from the decompactification 
of 3-dimensional non-supersymmetric models, but not supersymmetric models.

\section{An Explicit String Model}\label{Model}

{}In this section, we use an explicit 4-dimensional 
chiral ${\cal N}=1$ supersymmetric Type I string model as an 
illustration of some of the ideas discussed above.
Toroidal compactification of Type I string theory 
on a six dimensional torus $T^6$ gives rise to a four dimensional model with
${\cal N}=4$ supersymmetry. One can reduce the number of
supersymmetries to ${\cal N}=1$ by orbifolding. 
For example, take $T^6=T^2 \otimes T^2 \otimes T^2$, where
each of the $T^2$
has a ${\bf Z}_3$ and a ${\bf Z}_2$ rotational symmetry.
The ${\bf Z}_3$ generator $g$ and the ${\bf Z}_2$ generator $R$ acts
on the complex coordinates $z_1,z_2,z_3$
of the compactified dimensions as follows:
\begin{eqnarray}
g z_1 &=& \omega z_1~, \quad \quad g z_2 = \omega z_2, \quad \quad 
g z_3= \omega z_3 \\
R z_1 &=& - z_1~, \quad \quad R z_2 = - z_2, \quad \quad R z_3=  z_3
\end{eqnarray}
where $\omega=\exp(2 \pi i /3)$. The elements $g$ and $R$ generates
the group ${\bf Z}_6$. If we identify points in $T^6$ under this
discrete rotational symmetry, the resulting 
orbifold ${\cal M}=T^6/{\bf Z}_6$
has $SU(3)$ holonomy; only 1 of the 4 gravitinos are kept under the orbifold
action. As a result, Type I string theory compactified
on ${\cal M}$ has ${\cal N}=1$ supersymmetry in 4 dimensions.

{}To compute the spectrum, it is convenient to view Type I string
theory as Type IIB orientifold.
Type IIB string theory has a worldsheet reversal symmetry.
The orientifold projection $\Omega$
reverses the
parity of the closed string worldsheet (and hence interchanges
the role of left- and right-movers in Type IIB theory). 
Gauging this worldsheet parity symmetry
results in a theory of unoriented closed strings. Open strings and
$D$-branes are introduced to cancel the divergences (tadpoles) from the
Klein bottle amplitude (a one-loop amplitude for {\em unoriented} closed
strings). The orientifold group ${\cal O}$ (the discrete symmetries of 
Type IIB theory that we are gauging)
contains the elements $\Omega$ and $\Omega R$. Tadpole
cancelation requires introducing both $D9$- and
$D5$-branes. 
Global Chan-Paton charges associated with the D-branes
manifest themselves as gauge symmetry in space-time. As a result,
there are gauge fields from both $D9$- and $D5$-branes.

{}The details of the tadpole cancelation conditions 
and the construction of
${\bf Z}_6$ orientifolds can be found in appendix \ref{construction}.
First, consider the case where the untwisted 
NS--NS sector B-field background is zero; tadpole cancelation implies
that $n_9=n_5=32$, where $n_9$ ($n_5$) is the number of 9-branes (5-branes).
This means that the total rank of the gauge group 
(which comes from both 9-branes and 5-branes) 
is $32$. This model has gauge group 
$[SU(6) \otimes SU(6) \otimes SU(4) \otimes U(1)^3]^2$ and 
was first constructed
in Ref \cite{KS2}. Although the gauge group contains the standard
model gauge group $SU(3) \otimes SU(2) \otimes U(1)$, the
residual gauge symmetry is too large for the model to be phenomenologically
interesting.

{}In the presence of the untwisted NS--NS sector B-field background,
it was shown \cite{oldBij,Bij} that the rank of the gauge group
is reduced to $32/2^{b/2}$. Here, $b$ is the rank of the matrix $B_{ij}$
($i,j$ labels the complex coordinates of $T^6$).
Since we are compactifying Type I string theory on a 6 dimensional
manifold, $b=0,2,4,6$. 
The details of the construction of these models can be found in
Appendix \ref{construction}. For $b=2$, the model has
$[SU(2) \otimes SU(2) \otimes SU(4) \otimes U(1)^3]^2$ gauge symmetry,
which can be considered as a Pati-Salam like model with some extra 
global/gauge symmetry depending on the gauge coupling of the
9-brane and 5-brane gauge group. This is the model we
are going to study in more details in this paper.
For $b=4$, the gauge group is $[SU(2) \otimes SU(2) \otimes U(1)^2]^2$
which is too small to contain the standard model.
For $b=6$, the gauge group is $[SU(2) \otimes U(1)]^2$, again does not
contain the standard model.

{}Let us discuss in more details the spectrum of the model with 
$[SU(2) \otimes SU(2) \otimes SU(4) \otimes U(1)^3]^2$ gauge symmetry.
Open strings start and end on D-branes. 
Since there are two kinds of D-branes (9-branes and 5-branes), 
there are three types of open strings that we need to consider:
99, 55 and 59 open strings.
The open string spectrum of this model is given in Table \ref{Z6}.
Here, we consider all D5-branes sitting at the same orbifold fixed point.
The fact that the $99$ and the $55$ sector have the same spectrum follows
from T-duality. Since open string has only two end-points, the
charged matter fields are either bi-fundamentals or symmetric
(or anti-symmetric) representations of the gauge groups. 
Notice that the first and the second $U(1)$ of both the $99$
and the $55$ gauge groups are anomalous, with $U(1)$ anomaly equals
$-16$ and $+16$ respectively.
We can form a linear combination of these $U(1)$'s such that
only one of them is anomalous (this combination is given by
$Q_1 -Q_2$ where $Q_{1,2}$ are the first and the second $U(1)$ charge
respectively).
By the generalized Green-Schwarz mechanism \cite{GS}, some
of the fields charged under the anomalous $U(1)$ will acquire vevs
to cancel the Fayet-Illiopoulos $D$-term.
In addition to the open string spectrum, there are also
closed string states. Since they do not carry Chan-Paton factors, they
are singlets under the gauge group. 

{}We see that the model has enough realistic features so that we
can use it to study various scenarios discussed earlier. Here we shall
consider three different possible ways that the model may be interpreted as an
approximate way to describe nature. There is one chiral family in the
first scenario, two chiral families in the second scenario, 
and three chiral families in the third scenario.
Our description is sketchy and we shall
simply assume the dynamics needed to behave in the way
we like. Our purpose is to illustrate some of the features of brane-physics,
and draw attention to the model's
features that are generic to other Type I string models.
We shall not worry about which (if any) of the three scenarios is 
actually realized by the string dynamics.

\subsection{Scenario 1}

To describe this scenario, let us go to the T-dual picture where there
are two different types of 5-branes (as in Eq.(\ref{55action})). 
For convenience, 
we will still refer to the $5$-branes coming from T-dualizing
the $9$-branes as the $9$-branes.
Suppose the standard model $SU(3) \otimes SU(2) \otimes U(1)$
gauge group comes from the 9-brane
sector only. In this model, the gauge group is 
$SU(4) \otimes SU(2)_L \otimes SU(2)_R$, and the
99 sector matter fields are singlet under the 5-brane gauge group. We can make
the 5-brane gauge coupling relatively strong, so that $SU(4)_5$ gets strong and
may trigger dynamical supersymmetry breaking. It may also cause spontaneous
symmetry breaking of $SU(2)_5 \otimes SU(2)_5$ 
so that the 55 and the 59 sector matter
fields become heavy. In any case, let us focus our attention on the 99 sector.
Here some of the low dimension terms in the superpotential is given by
(see Table \ref{Z6} for notations)
\begin{equation}
{\cal W} = \left( U_1 Q_2 + U_2 Q_1 \right) H
              +\left( U_1 S_2 + U_2 S_1 \right) {U}_3
              +\left( Q_1 S_4 + Q_2 S_3 \right) {Q}_3 + \dots
\end{equation}
where we have suppressed the $\lambda$
dependence and the exact coefficients of the couplings. 
(The $\lambda$ dependence of $N$-point couplings
is $g^{N-2} \sim \lambda^{(N-2)/2}$).

To break the gauge group down to $SU(3) \otimes SU(2) \otimes U(1)$, 
we can move some of the 9-branes away from each other. 
This mechanism is equivalent to the spontaneous symmetry breaking (SSB) 
action of the Higgs field in the effective field theory; that is, 
we can give vacuum expectation value to 
the Higgs superpartner of one of the $U$ fields. 
Since the $U$ fields are charged under $U(4) \supset SU(4) \otimes U(1)$, 
it is more appropriate to 
consider $SU(4) \otimes SU(2)_L \otimes SU(2)_R \otimes U(1)
\supset
SU(3) \otimes SU(2)_L \otimes U(1) \otimes U(1) \otimes U(1)$,
\begin{equation}
(\overline{\bf 4},{\bf 1},{\bf 2})(-1) =
(\overline{\bf 3},{\bf 1})(\textstyle{-{1 \over 3},{1 \over 2},-1}) \oplus
(\overline{\bf 3},{\bf 1})(\textstyle{-{1 \over 3},-{1\over 2},-1}) \oplus
({\bf 1},{\bf 1})(\textstyle{1,{1 \over 2},-1}) \oplus
({\bf 1},{\bf 1})(\textstyle{1,-{1 \over 2},-1})
\end{equation}
Here, the first $U(1)$ charge is the $B-L$ number, the second $U(1)$
charge is $I_R= SU(2)_R$ isospin, and the third $U(1)$ charge is $3B+L$ which
comes from the decomposition
$U(4) \supset SU(4) \otimes U(1)$. Notice that the $U(1)$ hypercharge
$Y=B-L + 2 I_R$ and the baryon number $B= (B-L + 3 B + L)/4$.
Therefore, under $SU(4) \otimes SU(2) \otimes SU(2) \otimes U(1)
\supset SU(3) \otimes SU(2)_L \otimes U(1)_Y \otimes 
U(1)_B \otimes U(1)_{3B+L}$,
\begin{equation}
(\overline{\bf 4},{\bf 1},{\bf 2})(-1) =
(\overline{\bf 3},{\bf 1})(\textstyle{{2 \over 3},-{1\over 3},-1}) \oplus
(\overline{\bf 3},{\bf 1})(\textstyle{-{4 \over 3},-{1\over 3},-1}) \oplus
({\bf 1},{\bf 1})(\textstyle{2,0,-1}) \oplus
({\bf 1},{\bf 1})(\textstyle{0,0,-1})
\end{equation}
Here the $U(1)$s are independent but not orthogonal.
If the scalar $({\bf 1},{\bf 1})(0,0,-1)$ acquires a vev, $U(1)_{3B+L}$
is broken, and the fields $Q_i$ and $U_i$ become
\begin{eqnarray}
({\bf 4},{\bf 2},{\bf 1})(+1) &=& ({\bf 3},{\bf 2})
(\textstyle{{1\over 3},{1\over 3}}) \oplus
({\bf 1},{\bf 2})(-1,0) \nonumber \\
(\overline{\bf 4},{\bf 1},{\bf 2})(-1) &=&
(\overline{\bf 3},{\bf 1})(\textstyle{{2 \over 3},-{1\over 3}})
\oplus
(\overline{\bf 3},{\bf 1})
(\textstyle{-{4 \over 3},-{1\over 3}}) \oplus
({\bf 1},{\bf 1})(2,0) \oplus
({\bf 1},{\bf 1})(0,0) 
\end{eqnarray}
We see that the $Q_i$ and $U_i$ yield precisely one chiral and one vector 
({\em i.e.}, one chiral plus one anti-chiral) family of the standard model 
$SU(3) \otimes SU(2)_L \otimes U(1)_Y \otimes U(1)_B$.
This also splits the $SU(2)_R$ doublet $H$ into two standard models doublets 
$H_1$ and $H_2$:
\begin{equation}
({\bf 1},{\bf 2},{\bf 2})(0) =
({\bf 1},{\bf 2})(1,0) \oplus
({\bf 1},{\bf 2})(-1,0)
\end{equation}
The $\mu$ term $\mu H_1 H_2$ does not appear as lower order terms in
the superpotential. In this scenario where there are only 9-branes, 
$g_3=g_2=g_9$, and $g_Y=\sqrt{3 \over 5} g_9$ at the string scale.

{}Consider the chiral fermions in the 99 sector before the electroweak symmetry
breaking. There is 1 chiral family and 1 vector (chiral plus anti-chiral)
family. Generically, a linear combination of $U_1$ and $U_2$ 
will pair up with $U_3$ to become
heavy, while the other linear combination will remain massless.  
After the SSB
to $SU(3) \otimes SU(2) \otimes U(1)$, this 
$\alpha U_1 + \beta U_2$ combination gives the right-handed quarks
and leptons.
Similarly, a linear combination of $Q_1$ and $Q_2$ may pair up with 
$Q_3$ to become
heavy, while the other linear combination will remain massless.
They yield the weak isodoublets of quarks and leptons. So we see that the
model has only one chiral family of quarks and leptons.

Now, notice that there are no baryon number violating terms in the
superpotential. This is due the third $U(1)$ symmetry. The quarks have $U(1)_3$
charge +1, while the antiquarks have charge $-1$. The presence of such a $U(1)$
associated with the $SU(4)$ is a generic feature of brane physics (the $U(1)$
factor is the center of mass of the D-branes). So we should
expect the conservation of the baryon number as a generic feature.

Suppose, in Eq.(\ref{55action}), it is $v_1$, not $v_3$, that is becoming 
very large. In this case, the $5$-brane sector gauge coupling becomes 
vanishingly small. So the $5$-brane matter fields essentially decouple and 
can be candidates for dark matter.

\subsection{Scenario 2}

	Suppose the QCD $SU(3)$ comes from the 9-brane sector while the 
weak $SU(2)$
comes from the 5-brane sector. To be specific, the gauge group is
$SU(4)_9 \otimes SU(2)_5 \otimes SU(2)_5$. 
The quarks and leptons come from the 59 sector while
the Higgs field comes from the 55 sector. There is a ${\bf Z}_2$ 
symmetry under which
all matter fields are odd while the Higgs field is even. The superpotential is
given by
\begin{eqnarray}
{\cal W}=&& \left( U_1 Q_2 + U_2 Q_1 \right) H
              +\left( U_1 S_2 + U_2 S_1 \right) {U}_3
              +\left( Q_1 S_4 + Q_2 S_3 \right) {Q}_3 \nonumber \\
          &+& \left( u_1 q_2 + u_2 q_1 \right) h
              +\left( u_1 s_2 + u_2 s_1 \right) {u}_3
              +\left( q_1 s_4 + q_2 s_3 \right) {q}_3 \nonumber \\
          &+& \sum_{i=1}^2 \sum_{j=1}^2 
              {\cal U}_i {\cal Q}_j h + 
              \sum_{i=1}^2 \sum_{j=3}^4
              {\cal U}_i {\cal U}_j H 
             + \dots
\end{eqnarray}
Again, we have suppressed the $\lambda$ dependence and the exact coefficients
of the couplings. As before, vev for one of ${\cal U}_i$ fields induces SSB:
$SU(4) \otimes SU(2)_L \otimes SU(2)_R \supset SU(3) \otimes SU(2) \otimes
U(1)_Y$. There are two families of quarks and leptons.
As in the previous scenario, 
conservation of the third $U(1)$ charge 
prevents any perturbative baryon number
violating term. The analysis is quite similar to the above scenario, so we
shall not repeat. A crucial difference is that, even at the string scale, the
QCD coupling $g_3=g_9$ and the weak coupling $g_2=g_5$; they need 
not be the same. From Eq.(\ref{gaugecoupling}),
we see that their relative values depend on the compactification volumes.
The hypercharge $U(1)$ coupling is a function of $g_3$ and $g_2$:
\begin{equation}
g_Y = {{\sqrt{3} g_9 g_5} \over {\sqrt{{3 g_9^2 + 2 g_5^2}}}}
\end{equation}
If $g_9=g_5=g$, then $g_Y=\sqrt{3 \over 5} g$ at the string scale.

\subsection{Scenario 3}

We see that the model has 1 chiral family in the 99 sector and 
2 chiral families in the 59 sector. Furthermore, there is a 
${\bf Z}_2$ symmetry between the 9-brane and the 5-brane. 
We can construct a new model by gauging this ${\bf Z}_2$
symmetry, or part of it, i.e., a ${\bf Z}_2$ orbifold of the original model. 
The ${\bf Z}_2$ symmetry we want to orbifold is an outer-automorphism. 
In terms of current algebra in conformal field theory, such an orbifold 
converts level-1 current algebra to level-2 current algebra. 

{}Similar procedures can be carried out in the effective field theory
without having to impose the condition that $g_9=g_5$ \cite{barbieri}.
The basic idea is as follows. We start from a product gauge group
$SU(N) \otimes SU(N)$, with gauge couplings $g^{\prime}$ and 
$g^{\prime \prime}$ respectively.
By giving vev to the bi-fundamental field
$\phi=({\bf N},{\bf N})$ along the flat direction 
$\langle \phi \rangle = v I_N $
(where $I_N$ is an $N \times N$ identity matrix), 
the gauge group is broken to
$SU(N)$.
% with gauge coupling 
%\begin{equation}
%g = {{g^{\prime} g^{\prime\prime}} \over {\sqrt{(g^{\prime})^2 + 
%(g^{\prime \prime})^2}}}
%\end{equation}

{}In the specific model that we consider in this paper,
the fields
${\phi}_1$, ${\phi}_2$ are bi-fundamentals under the 
$U(2)_9 \otimes U(2)_5$ gauge group. Similarly, $\phi^{\prime}_1$, 
$\phi^{\prime}_2$ are 
bi-fundamentals under $U(2)^{\prime}_9 \otimes U(2)_5^{\prime}$.
By giving vevs to $\phi_i$'s and $\phi_i^{\prime}$'s 
of the above form (with $N=2$):
\begin{eqnarray}
U(2)_9 \otimes U(2)_5 &\rightarrow& SU(2)_L \otimes U(1) 
\nonumber \\
U(2)^{\prime}_9 \otimes U(2)^{\prime}_5 &\rightarrow& SU(2)_R \otimes U(1) 
\end{eqnarray}
The gauge couplings of $SU(2)_L$, $SU(2)_R$ and the accompanying
$U(1)$'s are given by
$g=g_9 g_5 / \sqrt{g_9^2+g_5^2}$. 
The $U(1)s$ are broken by the Green-Schwarz mechanism, so 
the resulting model has Pati-Salam gauge group 
$SU(4)_9 \otimes SU(2)_L  \otimes SU(2)_R$ with 
additional custodial $SU(4)_5 \otimes U(1)^2$
symmetry. There are three families of chiral fermions 
under the Pati-Salam gauge group. Two of them come
from the $59$ sector: ${\cal U}_i$ give rise to two families of
right-handed quarks and
leptons, while ${\cal Q}_i$ give rise to two families of left-handed
quarks and leptons. The remaining family comes from the $99$ sector:
the right-handed quarks and leptons come from a linear combination of
$U_1$ and $U_2$, and the left-handed quarks and leptons come from
a linear combination of $Q_1$ and $Q_2$. It is interesting to note
that one of the three families has a different origin. Whether
this will offer an explanation to the fact that there is one heavy family
deserves further investigation. Note that weak interaction universality 
is automatic.

{} The SSB of $SU(4) \otimes SU(2)_L \otimes SU(2)_R \supset SU(3) \otimes 
SU(2)_L \otimes U(1)_Y$ is essentially the same as in the first scenario,
that is, giving a vev to the one of the U fields. 
The gauge couplings of the standard model
gauge groups do not need to meet at the string scale and are given by
\begin{eqnarray}
g_3 &=& g_9 \nonumber \\
g_2 &=& {{g_9 g_5} \over \sqrt{g_9^2 + g_5^2}} \\
g_Y &=& {{\sqrt{3} g_3 g_2} \over \sqrt{3 g_3^2 + 2 g_2^2}} 
   \nonumber
\end{eqnarray} 
so that $\sin^2 \theta_W$ is given by
\begin{equation}
\sin^2 \theta_W = {{3 g_3^2} \over {6 g_3^2 + 2 g_2^2}}
\end{equation}
If $g_9=g_5=g$, we see that $g_3=g$, $g_2=g/\sqrt{2}$, $g_Y=\sqrt{3\over 8}
g$ and $\sin^2 \theta_W = 3/7$ at the string scale.
 
{}What about the $U(1)_B$ gauge boson associated with the baryon number 
conservation? Even if its coupling is very weak, it certainly must 
pick up a mass for the model to be phenomenologically viable. It is easy 
to see that this is possible only if QCD $SU(3)$ is broken as well, which 
implies that free quarks and gluons can exist. Suppose
the $U(1)_B$ boson picks up a mass $\mu$, then, 
following Ref \cite{giles}, there are free quarks and gluons with mass
about $(1~{\mbox{GeV}})^2/ {\mu}$. For $\mu=$ 10 keV, we see that a free quark 
or a free gluon will have a mass around 100 TeV.

\subsection{Another String Model}

{}Let us consider another ${\cal N}=1$, $D=4$ chiral Type I model, namely the
${\bf Z}_3 \otimes {\bf Z}_2 \otimes {\bf Z}_2$ model recently constructed by 
Kakushadze \cite{zura6}. This model has 9-branes and three
types of 5-branes as given in Eq.(\ref{95action}),
all of them have identical gauge groups,
so the resulting gauge group is $[U(6) \otimes SO(5)]^4$.
Let us assume that QCD $SU(3)$ comes from one of the $U(6)$,
while the weak $SU(2)$ comes from one of the $SO(5)$.
It seems there are enough Higgs fields to break one of the $SU(6)$ down 
to $SU(4)$ and then to $SU(3)$, and one of the $SO(5)$ to $SU(2)$.
Again, we see that the $U(1)$ carrying baryon numbers is present.
However, this $U(1)$ is anomalous, so it will pick up a mass via the 
Green-Schwarz mechanism automatically.
Consider the situation where the torus $T_3$ is very large. Following
Eq.(\ref{95action}), we see that both the gauge couplings of the $9$-brane
and the third $5$-brane sectors become vanishingly small. In particular,
the $9$-brane matter fields essentially decouple and can be candidates for
dark matter.

\section{Discussion}

{}It is clear that, among other properties, perturbative $D=4$, ${\cal N}=1$
supersymmetric, chiral Type I string models have some very attractive 
features for the study of the TeV scale string scenario : \\
(i) Gravitons live in the bulk while gauge and charged matter fields 
live on the branes. \\
(ii) The presence of $U(1)$s (associated with the centers of mass of the 
branes) which help to stabilize the proton. \\
(iii) The identical nature of $9$-brane gauge group and $5$-brane gauge 
groups (if present) allows different standard model gauge couplings
at the string scale. \\
(iv) the presence of bi-fundamental matter fields allows diagonal 
spontaneous symmetry breaking; again this mechanism allows different 
standard model gauge couplings at the string scale. This feature may 
validate the weak string coupling description of Type I string. \\ 
These properties are quite compatible with present experiments and allow 
the future tests of the extra large dimension scenario. 

{}There are a number of reasons why this TeV scale superstring scenario was
not seriously considered earlier. In the old string phenomenology framework,
({\em i.e.}, pre-string-duality days),
gravity and gauge interactions live in the same space. Since gauge
interactions clearly live in an effective 4 spacetime dimensions, at least up
to the electroweak scale, the largest the extra dimensions can be
is $M_{EW}^{-1}$, as considered in \cite{Antoniadis}.
However, generically, the string scale is above $M_{GUT}$ to satisfy
the proton decay bound.
The reason is following. Before our understanding of string duality, all
phenomenologically interesting string models are within the heterotic string
theory in the conformal field theory framework, where the original rank of the
gauge group is 22. Although the rank of the massless gauge symmetry can be
substantially reduced, the massive sector retains (at least some of) the
original large group feature. A typical heterotic string model that contains
the standard model of strong and electroweak interactions in its low energy
sector will contain massive bosons that can mediate proton decay. Since these
massive bosons have masses of the string scale, we must keep the string scale
high enough, say around $M_{GUT}$, to satisfy the proton decay bound.
Generically, the proton decay bound requires the absence of dimension-4
and -5 baryon-number violating operators.
If the string scale is around 1 TeV, the higher-dimensional (up to
dimension-18) baryon-number violating operator terms can be dangerous.
To prevent their appearance, some discrete symmetry or custodial gauge 
symmetry is necessary. However, the presence of such symmetry is not generic
in the old heterotic string theory.
In comparison, the $U(1)$'s in Type I strings are very generic; they 
correspond to the center of mass of the D-branes.
As we have seen in some cases, the difficulty is how to make them massive. 

{}Suppose we consider the heterotic string beyond the world-sheet construction.
For example, solitonic 5-branes can contribute to the massless spectrum 
in non-perturbative
heterotic string, which may have properties that are suitable for
phenomenology.  However, the analysis of non-perturbative heterotic string is
difficult. Hopefully, duality between the Type I and the heterotic 
string \cite{PW}
allows us to treat more fully the non-perturbative effects.

{}The string model that we have presented here is constructed from
perturbative Type I string theory. If the gauge coupling is of 
order 1, and we expect $m_s R > 1$, Eq.(\ref{55action}) implies that the 
string coupling $\lambda$ is small.
%\sim (m_s r)^2$. Clearly, $m_s r > 1$, and for 
%$m_s r \sim 10$, $\lambda \sim 100$. 
One would still like to know the energy regime where 
the perturbative Type I picture may become invalid \cite{kaplunovsky}.
Naively, one may expect the $4$-dimensional low energy effective 
field theory to be valid at momentum scales below $r^{-1}$. This is 
because the 
low energy effective couplings are small (except for the strong QCD coupling).
Quantum corrections coming from the massive string modes are negligible at low 
energies. Above this scale, one 
expects the $(4+n)$-dimensional effective field theory to be valid. At 
scales above $R^{-1}$ but below $m_s$, we should move from effective field 
theory to string theory, where perturbative Type I string theory is 
likely to be valid. When the energy-momentum scale is around the string 
scale $m_s$, 
the Type I string perturbative description may or may not remain valid.
This may depend on the particular scenario and the particular process one 
is interested in.
In view of Type I--heterotic duality, 
one would ask if the weakly-coupled heterotic
string description should take over in this regime.
However, the techniques in constructing 
heterotic string vacua with NS 5-branes
(the NS 5-branes are dual to the $D5$-branes in the Type I theory)
are not well developed. Since Type I string theory provides a natural setting
to realize the idea of extra large dimensions, it is likely that the
scenarios that we presented here capture the important features which
persist in the large $\lambda$ regime.

\section{Comments}

{}It is interesting to compare the merits of the two scenarios of string
phenomenology: the old scenario with string scale around the GUT scale, and
the new scenario \cite{lykken,extra,dienes} 
with the string scale around the electroweak scale.
Experimentally, the new scenario is clearly superior. High energy scatterings
can probe the extra small dimensions while gravity can probe the extra large
dimensions. These experiments are coming in the near future. If this scenario
is correct, we can expect a lot of experimental information on the detailed
structure, which can provide valuable guidance on the precise way nature is
realized within string theory.
At this moment, before the availability of the experimental data, we can still
ask which scenario is more appealing from the theoretical perspective. Without
detailed realistic models, any comparison is quite subjective. Nevertheless,
we believe the exercise can be illuminating. A scenario may be deemed more 
natural than another if it has fewer number of disparate scales. 
Let us give a naive counting of the number of scales in each scenario.

{}In the old scenario where the string scale is around the Planck scale 
$M_P$, we also
have the electroweak scale. The Planck scale $M_P$ is about three orders of
magnitude above the GUT scale; this discrepancy is different enough to 
require some new
physics ingredients to explain. Let us count this situation as 3 scales.
The quark and lepton masses are very different. For example, the mass of the
top quark is more than $10^5$ that of the electron. Let us assume that the
fermion mass splitting introduces another scale that needs understanding.
Including the cosmological constant, we have 5 different scales. Let us take
one of them, say the electroweak scale, to set the overall normalization.
Unification of the gauge couplings provides a nice explanation of the GUT
scale, so there remains 3 scales that remain to be understood. If one wants to
treat the GUT scale and the Planck scale as close enough to be considered as
one, we still have two scales that beg for an explanation.
 
{}In the new scenario, we have the string scale around 1 TeV, which is close
enough to the electroweak scale to be considered as a single scale. Similarly,
the small compactification radii between the electroweak and the
string scale should not be treated as new scales. Suppose the
standard model gauge couplings are unified at the string scale. Since the
gauge and matter fields are living in extra dimensions, say 8 total spacetime
dimensions, the gauge couplings are irrelevant operators. So these couplings
run as powers and diverge rapidly as we move to lower energies. Once the
energies involved go below the scale of the small radii, they become marginal
operators and vary only logarithmically. Suppose the Yukawa couplings at the
string scale are different but comparable. Again, as irrelevant operators,
they diverge rapidly, so they can easily differ by orders of magnitude at
scales below the electroweak scale \cite{dienes}. This provides a 
qualitative explanation for the fermion mass hierarchy. So we shall not 
count the fermion mass splittings as an extra scale.
Now we can count the number of scales
in this scenario : using the string/electroweak scale to set the overall
normalization, we have only two 
scales that beg for an explanation:
the cosmological constant and the large radius of $r=1$ mm$=10^{16}$/TeV. 
(In fact,
a cosmological constant of the order $r^{-4}$ is quite compatible with
observations.)  

{}Theoretically, it seems that the new scenario looks slightly better than, or
at least comparable to, the old scenario. Experimentally, the new scenario is
much more testable/reachable and hence superior. So overall, the new scenario
certainly deserves further investigation.  

\acknowledgments

{}We would like to thank Philip Argyres, Chong-Sun Chu, Keith Dienes, 
Alon Faraggi, Piljin Yi, and especially Zurab Kakushadze 
for valuable discussions. 
The research of G.S. and S.-H.H.T. was partially supported by the 
National Science Foundation. 
G.S. would like to thank the kind hospitality of the Institute for
Theoretical Physics at Stony Brook during his stay.
G.S. would also like to thank
Joyce M. Kuok Foundation for financial support.

\appendix
\section{Construction of the Model}\label{construction}

{}In this appendix, we give the details of how to construct from D-branes
and orientifolds
the ${\cal N}=1$, $D=4$ chiral string model with 
$[SU(4) \otimes SU(2) \otimes SU(2) \otimes U(1)^3]^2$ gauge symmetry
presented in Section \ref{Model}.
The model also exhibits some novel features \cite{oldBij}
of the untwisted
NS--NS sector B-field background
recently discussed in \cite{Bij}.

{}We start from Type IIB string theory compactified on 
$T^6=T^2 \otimes T^2 \otimes T^2$, where
each of the two-tori has a ${\bf Z}_3$ and a ${\bf Z}_2$ rotational
symmetry. The ${\bf Z}_3$ generator $g$ and the ${\bf Z}_2$ generator $R$ 
acts on the complex coordinates $z_1,z_2,z_3$ of $T^6$
as follows:
\begin{eqnarray}
g z_1 &=& \omega z_1~, \quad \quad g z_2 = \omega z_2, \quad \quad 
g z_3= \omega z_3 \\
R z_1 &=& - z_1~, \quad \quad R z_2 = - z_2, \quad \quad R z_3=  z_3
\end{eqnarray}
where $\omega=\exp(2 \pi i /3)$.
The elements $g$ and $R$ together 
generate the Abelian group ${\bf Z}_6$.

{}Let us consider Type I string theory compactified on the toroidal
orbifold ${\cal M}= T^6/ {\bf Z}_6$. It is convenient to view Type I
compactification as Type IIB orientifold.
The orientifold projection $\Omega$
reverses the parity of the closed string worldsheet. 
This results in a theory of
unoriented closed strings. One-loop finiteness
generically requires introducing open strings starting and ending
on D-branes, so that the divergences (tadpoles) coming from the cylinder,
Mobius strip and Klein bottle amplitudes cancel. 
The orientifold group 
${\cal O}=\{ \Omega^a R^b g^c ~\vert ~ a=0,1~;b=0,1 ~; c=0,1,2 \}$ 
contains both the elements $\Omega$ and $\Omega R$. Therefore, 
one has to introduce both $D9$- and $D5$-branes
to cancel the tadpoles. The global Chan-Paton charges associated with
the D-branes manifest themselves as gauge symmetries in space-time.
Hence, there are gauge bosons from both $99$ and $55$ open strings.

{}The orbifold action on the Chan-Paton factors is described by unitary
matrices $\gamma_{k,p}$ that acts on the string end-points ($k$ labels the
orbifold group element, $p$ labels the type of branes).
Let $\vert \psi,ij \rangle$ be an open string state, where $\psi$ is
the state of the worldsheet fields and $i$, $j$ are the Chan-Paton
factors of the string end-points (the open string starts on a 
$p$-brane and ends on a $q$-brane).
The action of the orbifold element
$k$ is given by
\begin{equation}
k: \quad \quad \vert \psi,ij \rangle  \rightarrow
( \gamma_{k,p} )_{i i^{\prime}}
\vert k \cdot \psi,i^{\prime} j^{\prime} \rangle
( \gamma_{k,q}^{-1})_{j^{\prime} j}
\end{equation}
Tadpole cancelation determines the form of the $\gamma_{k,p}$ matrices.
There are two types of constraints that we need to consider. The first one
comes from the cancelation of the untwisted tadpoles for the $D9$-branes
and the $D5$-branes respectively. 
This type of constraint determines the number of $D9$- and $D5$-branes. In the
general case where the untwisted NS--NS sector B-field
can be non-vanishing (with $b$ equals 
the rank of the matrix $B_{ij}$, which is always even), 
tadpole cancelation for the untwisted 
R-R 10-form potential gives \cite{oldBij,Bij}
\begin{equation}
2^{b} ({\mbox {Tr}} (\gamma_{1,9}))^2 - 2^{b/2} 64
{\mbox {Tr}} (\gamma_{1,9}) + (32)^2 = 0~.
\end{equation}
Therefore, the number of D9-branes is given by $n_9=32/2^{b/2}$.
Similarly, tadpole cancelation condition
for the untwisted R-R 6-form potential gives \cite{Bij}
\begin{equation}
({\mbox {Tr}} (\gamma_{1,5}))^2 - {64 \over 2^{b/2}} {\mbox {Tr}} 
(\gamma_{1,5}) 
+ {1\over 2^b} (32)^2 = 0 ~.
\end{equation}
Therefore, the number of D5-branes is given by $n_5=32/2^{b/2}$. 
This was also expected from T-duality between D9- and D5-branes.

{}The other constraint 
comes from tadpole cancelation of the twisted R-R 6-form potential.
Since the twisted closed string states propagating in the
tree-channel do not have momentum or winding, the twisted tadpoles
remain the same in the presence of the untwisted NS--NS sector B-field 
background (the effect of which
is to shift the left- plus right-moving momentum lattice).
The twisted tadpoles for ${\bf Z}_N$ orientifolds in 6 dimensions 
have been computed in
\cite{PS,GP,GJ} and generalized to 4 dimensions in 
Ref \cite{Sagnotti,KS1,KS2}. 
Here, we state the results for the ${\bf Z}_6$ case:
\begin{eqnarray}
{\mbox Tr} \left( \gamma_{g,p} \right) &=& -(-1)^{b/2} 32 
\left [\cos ( \pi/3) \right]^3
=-(-1)^{b/2} 4  \\
{\mbox Tr} \left( \gamma_{R,p} \right) &=& {\mbox Tr} \left( \gamma_{Rg,p} 
\right) = 0
\end{eqnarray}
 
{}Let us consider the solutions to the above tadpole cancelation conditions
for all possible values of $b$:

\noindent $\bullet$ For $b=0$, $n_9=n_5=32$
\begin{eqnarray}
\gamma_{R,p} &=& {\mbox {diag}} 
\left( i I_{16}, -i I_{16}  \right)~,  \\ 
\gamma_{g,p} &=& {\mbox{diag}} \left( \omega I_6,
\omega^2 I_6, I_4,\omega I_6,
\omega^2 I_6, I_4 \right)~. 
\end{eqnarray}
where $I_M$ is an $M \times M$ identity matrix.
The gauge group from the $99$ open strings
is $SU(6) \otimes SU(6) \otimes SU(4) \otimes U(1)^3$. 
The $55$ open strings also give rise to the gauge group
$SU(6) \otimes SU(6) \otimes SU(4) \otimes U(1)^3$ if the $D5$-branes
are located at the same fixed point. The total rank of the gauge group is
$32$. This model was first constructed in
Ref \cite{KS2}.

\noindent $\bullet$ For $b=2$, $n_9=n_5=16$.
\begin{eqnarray}
\gamma_{R,p} &=& {\mbox {diag}} 
\left( i I_8, -i I_8  \right)~,  \\ 
\gamma_{g,p} &=& {\mbox{diag}} \left( \omega I_2 ,
\omega^2 I_2, I_4, \omega I_2,
\omega^2 I_2, I_4 \right)~. 
\end{eqnarray}
The gauge group (from both $99$ and $55$ open strings)
is $[SU(2) \otimes SU(2) \otimes SU(4) \otimes U(1)^3]^2$. 
The total rank of the gauge group is $16$. 
This is the model that we study in this paper.

\noindent $\bullet$ For $b=4$, $n_9=n_5=8$.
\begin{eqnarray}
\gamma_{R,p} &=& {\mbox {diag}} 
\left( i I_4, -i I_4  \right)~,  \\ 
\gamma_{g,p} &=& {\mbox{diag}} \left( \omega I_2,
\omega^2 I_2,
\omega I_2,
\omega^2 I_2 \right)~. 
\end{eqnarray}
The gauge group (from both $99$ and $55$ open strings)
is $[SU(2) \otimes SU(2) \otimes U(1)^2]^2$. 
The total rank of the gauge group is $8$. 

\noindent $\bullet$ For $b=6$, $n_9=n_5=4$.
\begin{eqnarray}
\gamma_{R,p} &=& {\mbox {diag}} 
\left( i I_2, -i I_2  \right)~,  \\ 
\gamma_{g,p} &=& I_4 ~.
\end{eqnarray}
The gauge group (from both $99$ and $55$ open strings)
is $[SU(2) \otimes U(1)]^2$. 
The total rank of the gauge group is $4$. 

{}The gauge groups of the models for $b=4,6$ are too small to 
accommodate the standard model, which make them phenomenologically
uninteresting.  We will focus on the $b=2$ model with 
$[SU(2) \otimes SU(2) \otimes SU(4) \otimes U(1)^3]^2$ gauge symmetry.

{}To construct the open string spectrum, we keep all physical states
that are invariant under the orbifold action. There are contributions 
from $99$, $55$ and $59$ open strings. As pointed out in Ref \cite{Bij},
the $59$ open string sector states come with a multiplicity $\xi=2^{b/2}$.
(Recall that without B-field, 
the multiplicity of states in the $59$ sector was one per 
configuration of Chan-Paton charges \cite{GP,GJ}).
The open string spectrum of the 
$[SU(2) \otimes SU(2) \otimes SU(4) \otimes U(1)^3]^2$ model
is given in Table \ref{Z6}.

\section{$H$-charges, Scatterings and Couplings}

{}In this appendix, we review the conformal field theory
techniques in calculating
scattering amplitudes (and hence couplings) in orbifold models.
In Type I string theory, closed string sector only gives rise to 
gauge singlets.
We will therefore focus on the couplings between open string states.

{}In the standard orbifold formalism, the internal part of the 
worldsheet supercurrent can be written as
\begin{equation}
 T_F={i\over 2}\sum_{a=1}^{3} \psi^a \partial X^a +{\mbox {H.c.}}
={i \over 2} \sum_{a=1}^{3} e^{i \rho^{a}} 
                          \partial X^a + {\mbox {H.c.}} ~,
\end{equation}
where $\psi^a$ are complex world-sheet fermions, which can be bosonized:
\begin{eqnarray}
\psi^{a}&=& \exp(i \rho^{a}) =\exp(i H \cdot \rho) ~, \nonumber \\
\psi^{a \dagger}&=& \exp(-i \rho^{a}) =\exp(-i H \cdot \rho) ~.
\end{eqnarray}
Here, $H$ (known as the $H$-charge) equals $(1,0,0)$, $(0,1,0)$ or $(0,0,1)$ 
for $a=1,2,3$.
The supercurrent is therefore a linear combination of terms with 
well defined $H$-charges. 

{}In the covariant gauge, we have the
reparametrization ghosts $b$ and $c$, and superconformal ghosts
$\beta$ and $\gamma$ \cite{FMS}. It is most convenient to
bosonize the $\beta,\gamma$
ghosts:
$\beta = \partial \xi e^{-\phi}$, $\gamma = \eta e^{ \phi}$,
where $\xi$ and $\eta$ are auxiliary fermions and $\phi$ is a bosonic
ghost field obeying the OPE
$\phi(z) \phi(w) \sim
{\mbox{log}} ( z - w )$. The conformal dimension of
$e^{q \phi}$
is $-{1\over 2} q (q+2)$.
In covariant gauge, vertex operators
are of the form $V(z) \lambda_{ij}$,
where $V(z)$ is a dimension $1$ operator
constructed from the conformal fields (which include the longitudinal
components as well as the ghosts), and $\lambda_{ij}$ is the Chan-Paton
wavefunction.
The vertex operators for space-time
bosons carry integral ghost charges ($q \in {\bf Z}$) whereas
for space-time fermions the ghost charges are half-integral
($q \in {\bf Z} + {1\over 2}$). Here, $q$ specifies the picture.
The canonical choice is
$q=-1$ for space-time bosons and $q=-{1\over 2}$ for space-time
fermions. We will denote the corresponding vertex operators by
$V_{-1} (z)$ and $V_{-{1\over 2}}(z)$,
respectively. Vertex operators in the $q=0$ picture (with zero ghost charge)
is given by {\em picture changing}~:
\begin{equation}
V_{0}(z)= \lim_{w \rightarrow z}{ e^{\phi} T_F (z)
 V_{-1}(w)}~.
\end{equation}

{}One can see that besides the supercurrent, 
open string states also carry $H$-charges.
The vertex operator for gauge bosons in the $-1$ picture 
is given by $\psi^{\mu} \lambda_{ij}$
where $\mu$ is the spacetime index. Therefore, they do not carry $H$-charges.
On the other hand, the vertex operator for matter fields in $99$ 
and $55$ sector is given by $\psi^a \lambda_{ij}$. Hence, in the $-1$ picture,
$H=(1,0,0)$, $(0,1,0)$ or $(0,0,1)$ depending
on which worldsheet fermion is excited. The moding of the worldsheet fermions
in the $59$ sector is different from that in the $99$ sector. Therefore,
in the $-1$ picture, matter fields in the $59$ sector
carry half-integral $H$-charges instead of integral $H$-charges. 
The $H$-charges of the massless
fields of the $[SU(4) \otimes SU(2) \otimes SU(2) \otimes U(1)^3]^2$ 
model is given in Table \ref{Z6}.
 
{}Having constructed the vertex operators for the massless states, one
can in principle compute the scattering amplitudes, or the corresponding
couplings in the superpotential. The coupling of $M$ chiral superfields in
the superpotential is given by the scattering amplitude of the component
fields in the limit when all the external momenta are zero.
Due to holomorphicity, one needs to consider only the scatterings of 
left-handed space-time fermions, with vertices $V_{-1/2}(z)$,
and their space-time superpartners.
Since the total $\phi$ ghost charge in any tree-level correlation 
function is $-2$, it is convenient to choose two of the vertex operators in 
the $-1/2$-picture,
one in the $-1$-picture, and the rest in the $0$-picture.
Using the $SL(2,{\bf C})$ invariance, the scattering amplitude is therefore
\begin{eqnarray}\label{scattem}
 {\cal A}_{M} =&& g^{M-2}_{\mbox{st}}~
{\mbox{Tr}} \left( \lambda^1 \lambda^2 \cdots \lambda^M \right)   
\nonumber \\
 && \times \int dz_{4}
  \cdots dz_{M} 
        \langle V_{-{1\over 2}}(0)V_{-{1\over 2}}(1)
          V_{-1}(\infty) V_{0}(z_4)
          \cdots V_{0}(z_M) \rangle ~,
\end{eqnarray}
where we have normalized the $c$ ghost part of the correlation function
$\langle c(0) c(1) c(\infty) \rangle$ to $1$.
To obtain the open string scattering amplitudes, we have to take the
integration variables $z_i$ to the real axis, with $z_i > z_{i+1}$. Now the
terms in the superpotential can be read off directly from the resulting
scattering amplitudes.
For a non-zero coupling, the sum of the $H$-charges 
%as well as the sum of the $Q$-charges
must be zero in the 
corresponding scattering 
amplitude. Note that the supercurrent carries terms 
with different $H$-charges. 
%and $Q$-charges. 
Because of picture
changing, $H$-charges
%and $Q$- charges 
are not global charges even though
they must be conserved exactly. 
In additional to the $H$-charge conservation, there is also a discrete
${\bf Z}_2$ symmetry coming from the orbifold twist. For the couplings
to be non-zero, the total twist in the scattering amplitude (\ref{scattem})
must be an integer. 
 
%%%%%%%%%%%%%Table I %%%%%%%%
%%%%%%%%%%%%%%%%%%%%%%%%%%%%%%%%%%%%%%%%%%%%%%%%%%%%%%%%%%%%%%%%%%%%%%%%%%%%%%%
\begin{table}[t]
\begin{tabular}{|c|c|l|l|l|}
%%%%%%%%%%%%%%%%%%%%%%%%%%%%%%%%%%%%%%%%%%%%%%%%%%%%%%%%%%%%%%%%%%%%%%%%%%%%
 Sector & Field
        & $[SU(2)^{\prime} \otimes SU(2)\otimes SU(4)\otimes U(1)^3]^2$
        & $(H_1,H_2,H_3)_{-1}$ & $(H_1,H_2,H_3)_{-1/2}$ \\
\hline
%%%%%%%%%%%%%%%%%%%%%%%%%%%%%%%%%%%%%%%%%%%%%%%%%%%%%%%%%%%%%%%%%%%%%%%%%%%%
%Closed & & &\\
%Untwisted & $5({\bf 1}, {\bf 1}, {\bf 1}; {\bf 1}, {\bf 1}, {\bf 1})
%(0,0,0;0,0,0)_L$  & & \\
%\hline
%%%%%%%%%%%%%%%%%%%%%%%%%%%%%%%%%%%%%%%%%%%%%%%%%%%%%%%%%%%%%%%%%%%%%%%%%%%%
%Closed &  & & \\
%${\bf Z}_3$ Twisted  & $15({\bf 1}, {\bf 1}, {\bf 1}; {\bf 1}, {\bf 1}, 
%{\bf 1})
%(0,0,0;0,0,0)_L$ & & \\
%\hline
%Closed & & &  \\
%${\bf Z}_6$ Twisted  & $3({\bf 1}, {\bf 1}, {\bf 1}; {\bf 1}, {\bf 1}, {\bf 1})
%(0,0,0;0,0,0)_L$ & & \\
%\hline
%Closed & & &  \\
%${\bf Z}_2$ Twisted  & $11({\bf 1}, {\bf 1}, {\bf 1}; {\bf 1}, {\bf 1}, 
%{\bf 1})
%(0,0,0;0,0,0)_L$ & & \\
%\hline
%%%%%%%%%%%%%%%%%%%%%%%%%%%%%%%%%%%%%%%%%%%%%%%%%%%%%%%%%%%%%%%%%%%%%%%%%%%%
& $S_1$           & $({\bf 3},{\bf 1},{\bf 1};{\bf 1},{\bf 1},{\bf 1})
(+2,0,0;0,0,0)_L$ & $(+1,0,0)$ & $(+{1\over 2},-{1\over 2},-{1\over 2})$ \\
& $S_2$          &  $({\bf 3},{\bf 1},{\bf 1};{\bf 1},{\bf 1},{\bf 1})
(+2,0,0;0,0,0)_L$  & $(0,+1,0)$ & $(-{1\over 2},+{1\over 2},-{1\over 2})$ \\   
& $S_3$          & $({\bf 1},{\bf 3}, {\bf 1}; {\bf 1},{\bf 1},{\bf 1})
(0,-2,0;0,0,0)_L$ & $(+1,0,0)$ & $(+{1\over 2},-{1\over 2},-{1\over 2})$ \\
& $S_4$          & $({\bf 1},{\bf 3}, {\bf 1}; {\bf 1},{\bf 1},{\bf 1})
(0,-2,0;0,0,0)_L$   & $(0,+1,0)$ & $(-{1\over 2},+{1\over 2},-{1\over 2})$ \\ 
& $U_1$          & $({\bf 2},{\bf 1},\overline{\bf 4};{\bf 1},{\bf 1},
{\bf 1})(-1,0,-1;0,0,0)_L$ & $(+1,0,0)$ 
& $(+{1\over 2},-{1\over 2},-{1\over 2})$ \\
Open $99$
& $U_2$          &   $({\bf 2},{\bf 1},\overline{\bf 4};{\bf 1},{\bf 1},
{\bf 1})(-1,0,-1;0,0,0)_L$ 
& $(0,+1,0)$ & $(-{1\over 2},+{1\over 2},-{1\over 2})$ \\ 
& $U_3$          
& $({\bf 2},{\bf 1},{\bf 4};{\bf 1},{\bf 1},{\bf 1})
(-1,0,+1;0,0,0)_L$ & $(0,0,+1)$ & $(-{1\over 2},-{1\over 2},+{1\over 2})$ \\
& $Q_1$ & $({\bf 1},{\bf 2},{\bf 4};{\bf 1},{\bf 1},{\bf 1})
(0,+1,+1;0,0,0)_L$ & $(+1,0,0)$ & $(+{1\over 2},-{1\over 2},-{1\over 2})$ \\
& $Q_2$         &   $({\bf 1},{\bf 2},{\bf 4};{\bf 1},{\bf 1},{\bf 1})
(0,+1,+1;0,0,0)_L$    & $(0,+1,0)$ & $(-{1\over 2},+{1\over 2},-{1\over 2})$ \\
& $Q_3$          & $({\bf 1},{\bf 2},\overline{\bf 4};{\bf 1},{\bf 1},{\bf 1})
(0,+1,-1;0,0,0)_L$ & $(0,0,+1)$ & $(-{1\over 2},-{1\over 2},+{1\over 2})$ \\  
& $H$           & $({\bf 2},{\bf 2},{\bf 1};{\bf 1},{\bf 1},{\bf 1})
(+1,-1,0;0,0,0)_L$ & $(0,0,+1)$ & $(-{1\over 2},-{1\over 2},+{1\over 2})$ \\
\hline
%%%%%%%%%%%%%%%%%%%%%%%%%%%%%%%%%%%%%%%%%%%%%%%%%%%%%%%%%%%%%%%%%%%%%%%%%%%
& $s_1$           & $({\bf 1},{\bf 1},{\bf 1};{\bf 3},{\bf 1},{\bf 1})
(0,0,0;+2,0,0)_L$ & $(+1,0,0)$ & $(+{1\over 2},-{1\over 2},-{1\over 2})$ \\
& $s_2$          &   $({\bf 1},{\bf 1},{\bf 1};{\bf 3},{\bf 1},{\bf 1})
(0,0,0;+2,0,0)_L$ & $(+1,0,0)$ & $(+{1\over 2},-{1\over 2},-{1\over 2})$\\ 
& $s_3$          & $({\bf 1},{\bf 1},{\bf 1};{\bf 1},{\bf 3},{\bf 1})
(0,0,0;0,-2,0)_L$ & $(+1,0,0)$ & $(+{1\over 2},-{1\over 2},-{1\over 2})$ \\
& $s_4$          &  $({\bf 1},{\bf 1},{\bf 1};{\bf 1},{\bf 3},{\bf 1})
(0,0,0;0,-2,0)_L$    & $(0,+1,0)$ & $(-{1\over 2},+{1\over 2},-{1\over 2})$ \\ 
& $u_1$         & $({\bf 1},{\bf 1},{\bf 1};{\bf 2},{\bf 1},
\overline{\bf 4})(0,0,0;-1,0,-1)_L$ & $(+1,0,0)$ 
                  & $(+{1\over 2},-{1\over 2},-{1\over 2})$ \\
Open $55$ & $u_2$          &  $({\bf 1},{\bf 1},{\bf 1};{\bf 2},{\bf 1},
\overline{\bf 4})(0,0,0;-1,0,-1)_L$    & $(0,+1,0)$ & $(-{1\over 2},+{1\over 2},-{1\over 2})$ \\
& $u_3$          
& $({\bf 1},{\bf 1},{\bf 1};{\bf 2},{\bf 1},{\bf 4})
(0,0,0;-1,0,+1)_L$ & $(0,0,+1)$ & $(-{1\over 2},-{1\over 2},+{1\over 2})$ \\
& $q_1$ &$({\bf 1},{\bf 1},{\bf 1};{\bf 1},{\bf 2},{\bf 4})
(0,0,0;0,+1,+1)_L$ & $(+1,0,0)$ & $(+{1\over 2},-{1\over 2},-{1\over 2})$ \\
& $q_2$          &   $({\bf 1},{\bf 1},{\bf 1};{\bf 1},{\bf 2},{\bf 4})
(0,0,0;0,+1,+1)_L$   & $(0,+1,0)$ & $(-{1\over 2},+{1\over 2},-{1\over 2})$ \\
 & $q_3$
          & $({\bf 1},{\bf 1},{\bf 1};{\bf 1},{\bf 2},\overline{\bf 4})
(0,0,0;0,+1,-1)_L$ & $(0,0,+1)$ & $(-{1\over 2},-{1\over 2},+{1\over 2})$ \\
& $h$          & $({\bf 1},{\bf 1},{\bf 1};{\bf 2},{\bf 2},{\bf 1})
(0,0,0;+1,-1,0)_L$ & $(0,0,+1)$ & $(-{1\over 2},-{1\over 2},+{1\over 2})$ \\
\hline
%%%%%%%%%%%%%%%%%%%%%%%%%%%%%%%%%%%%%%%%%%%%%%%%%%%%%%%%%%%%%%%%%%%%%%%%%%%
& ${\phi}_1^{\prime}$, ${\phi}_2^{\prime}$           
& $2({\bf 2},{\bf 1},{\bf 1};{\bf 2},{\bf 1},{\bf 1})
(+1,0,0;+1,0,0)_L$ & $(+{1\over 2},+{1\over 2},0)$ & $(0,0,-{1\over 2})$ \\
& ${\phi}_1$, ${\phi}_2$
         & $2({\bf 1},{\bf 2},{\bf 1};{\bf 1},{\bf 2},
{\bf 1})(0,-1,0;0,-1,0)_L$ 
& $(+{1\over 2},+{1\over 2},0)$ & $(0,0,-{1\over 2})$ \\ 
Open  $59$ & ${\cal Q}_1$, ${\cal Q}_2$
& $2({\bf 1},{\bf 1},{\bf 4};{\bf 1},{\bf 2},{\bf 1})
(0,0,+1;0,+1,0)_L$ & $(+{1\over 2},+{1\over 2},0)$ & $(0,0,-{1\over 2})$ \\
& ${\cal U}_1$, ${\cal U}_2$
         & $2({\bf 1},{\bf 1},\overline{\bf 4};{\bf 2},{\bf 1},
{\bf 1})(0,0,-1;-1,0,0)_L$
& $(+{1\over 2},+{1\over 2},0)$ & $(0,0,-{1\over 2})$ \\ 
& $\chi_1$, $\chi_2$
          & $2({\bf 1},{\bf 2},{\bf 1};{\bf 1},{\bf 1},{\bf 4})
(0,+1,0;0,0,+1)_L$ & $(+{1\over 2},+{1\over 2},0)$ & $(0,0,-{1\over 2})$ \\
& $\chi_3$, $\chi_4$
          & $2({\bf 2},{\bf 1},{\bf 1};{\bf 1},{\bf 1},
\overline{\bf 4})(-1,0,0;0,0,-1)_L$ 
& $(+{1\over 2},+{1\over 2},0)$ & $(0,0,-{1\over 2})$ \\ 
%%%%%%%%%%%%%%%%%%%%%%%%%%%%%%%%%%%%%%%%%%%%%%%%%%%%%%%%%%%%%%%%%%%%%%%%%%%
\end{tabular}
%%%%%%%%%%%%%%%%%%%%%%%%%%%%%%%%%%%%%%%%%%%%%%%%%%%%%%%%%%%%%%%%%%%%%%%%%%%
\caption{The massless open string 
spectrum of the 4-dimensional Type I ${\bf Z}_6$ orbifold model 
with $N=1$ space-time 
supersymmetry and gauge group 
$[SU(2)^{\prime} \otimes SU(2) \otimes SU(4) \otimes U(1)^3]^2$. 
The $U(1)$'s come from the traces of $U(2)^{\prime}$, $U(2)$ and 
$U(4)$ respectively. 
The $H$-charges in both the $-1$ picture and the $-1/2$ picture for states
in the open
string sector are also given. The vector multiplets are not shown.
The closed string sectors give rise
to the gauge singlets and the gravity supermultiplet. The $H$-charges
are explained in Appendix B.}  
\label{Z6}
\end{table}
%%%%%%%%%%%%%%%%%%%%%%%%%%%%%%%%%%%%%%%%%%%%%%%%%%%%%%%%%%%%%%%%%%%%%%%%%%%%%%%

\end{document}